\documentclass{nic-series}
\usepackage{epsf,psfig,epsfig}

\begin{document}

\title{Sequential Monte Carlo Methods for Protein Folding}

\author{Peter Grassberger }

\institute{Complex Systems Research Group, \\
           John-von-Neumann Institute for Computing\\
           Forschungszentrum J{\"u}lich, D-52425 J{\"u}lich, Germany}

\maketitle

\begin{abstracts}
We describe a class of growth algorithms for finding low energy states
of heteropolymers. These polymers form toy models for proteins, and 
the hope is that similar methods will ultimately be useful for finding
native states of real proteins from heuristic or a priori determined 
force fields. These algorithms share with standard Markov chain Monte
Carlo methods that they generate Gibbs-Boltzmann distributions, but 
they are not based on the strategy that this distribution is obtained 
as stationary state of a suitably constructed Markov chain. Rather, 
they are based on growing the polymer by successively adding 
individual particles, guiding the growth towards configurations with
lower energies, and using ``population control" to eliminate bad 
configurations and increase the number of ``good ones". This is not
done via a breadth-first implementation as in genetic algorithms, but
depth-first via recursive backtracking. As seen from various benchmark 
tests, the resulting algorithms are extremely efficient for lattice 
models, and are still competitive with other methods for simple 
off-lattice models.

\end{abstracts}

\section{Introduction}

Research in modern biology is more and more based on heavy numerical 
computations. This includes the reconstruction of complex structures
from X-ray imaging, sequence alignment, the optimization of phylogenetic 
trees, simulation of protein networks, and -- last but not least -- protein 
folding. The latter is considered by many as one of the most challenging
problems in mathematical biology. It is concerned with the problem of how
a heteropolymer of a given sequence of amino acids folds into precisely 
that geometric shape in which it performs its biological function 
\cite{Gierasch:King:90,Creighton:92,Merz:LeGrand:94,Brunak:95}.
Currently, it is much simpler to find coding DNA --- and, thus, also 
amino acid --- sequences than to elucidate the 3-$d$ structures of given
proteins. Therefore, solving the protein folding problem would be a 
major break-through in understanding the biochemistry of the cell.

Indeed, there are several closely related problems associated with
protein folding. What we have alluded to above is the {\it fold prediction}
problem. In addition there is the inverse fold prediction problem, 
concerned with designing a sequence which has a given 3-d structure as
its native state. This is obviously a central problem in ab initio 
drug design, and in designing artificial proteins. Again another problem
is that of discovering the detailed pathways followed during the folding
process. This is obviously important if one wants to interfere during 
it, as would be useful in treatments of cancer, BSE, and Alzheimer.

In the present review we shall only be concerned with the direct native 
state prediction problem. At present, the most successful methods 
are not based on physico-chemical principles like force fields
(potential energies), but on exploiting analogies with similar 
proteins whose structure has been determined by X-ray or NMR 
studies. This data-based approach will stay the method of choice 
for some time, but it cannot fully substitute a more fundamental
approach where one starts from a force field and identifies the native
state as the state with lowest free energy. The basis of such an 
approach are of the course the famous experiments of Anfinsen who
showed some thirty years ago that native states are uniquely defined
by their amino acid sequences, and that not too large proteins 
fold spontaneously and reversibly.

In such a direct ab initio approach there are two main difficulties.
The first is that presently available force fields are most likely
too crude for any but the most simple proteins. These fields are 
typically obtained by heuristic arguments and involve drastic
simplifications (such as using two-body potentials of Lennard-Jones
type, fixed bond lengths, etc.). Proteins typically are not very 
tightly collapsed (after all, most of them are catalyzers which 
have to be flexible in order to function), and the differences in 
energy between the native and misfolded states are typically as 
large or even smaller than the uncertainty of the potential energies.

The other problem is that algorithms for finding low energy states
of a complex potential are still too slow. There has been a dramatic
improvement during recent years in the efficiency of Monte Carlo
algorithms\cite{Okamoto,wille-hansmann,wang-landau,irbaeck,chikenji}, 
and presently available algorithms are fast enough to find the native 
states of small proteins (with up to 30 amino acids or more) in vacuum. 
But they are still not efficient for proteins in water (and simplified
treatments of the effect of water are problematic\cite{hsu-berg}),
so that the method of choice today still is molecular dynamics, 
even though it is by far too slow to study larger proteins. Notice
that fast minimization algorithms would also be extremely useful
in the search for better force fields.

\section{Chain growth methods}

In view of this situation, we shall discuss in the present review
the potential use of an alternative class of Monte Carlo methods
which are not based on the Metropolis concept of putting up a
Markov chain (with proposed moves and subsequent acceptance or 
rejection) whose stationary state is the wanted Gibbs-Boltzmann 
distribution. Instead, they are based on {\it sequential sampling}
or polymer {\it growth}. Starting with an empty configuration, 
particle after particle is added in a biased way to take into 
account steric and energetic effects. This bias is compensated
by a weight factor which multiplies the Boltzmann factor for the 
potential energy of the new particle in the field of the previous ones.
The product $W_n = \prod_{i=1}^nw_i$ of these factors (one for each newly 
added particle $i$) gives the total weight of a configuration 
containing $n$ particles. The power of such methods for 
generating valid configurations (not necessarily optimal ones) is 
discussed in Ref\cite{gan-schlick}.

If the bias is not perfect, i.e. if the bias correcting factor does
not compensate perfectly the Boltzmann factor, the total weight $W_n$
will fluctuate more and more as particles are added. As in genetic 
algorithms\cite{schwefel} and as in diffusion type quantum Monte Carlo
algorithms\cite{umrigar}, configurations with low weight are killed
while ``good", i.e. high weight configurations are split into two,
with the weight shared among them. In order to preserve the correct
statistics during this ``population control", the killing is not done
unconditionally but with probability 1/2: Low-weight configurations 
are randomly either killed or kept, in which case their weight is
doubled.

Algorithms of this type were first developed for neutron transport 
theory in the 1950's, when they were called ``Russian roulette and 
splitting"\cite{kahn}. For recent reviews see Ref\cite{liu,doucet_etal}. 
They were re-invented several times, and applied to proteins first 
by Velikson {\it et al.}\cite{velikson}. For a recent paper see 
Ref\cite{zhang02}. In all these works, the population control was made
``breadth first" as in genetic algorithms and in diffusion quantum
Monte Carlo: A large number (ca. 100 or more) of replicas are treated
simultaneously, and the ``goodness" of a configuration is determined 
by comparing it with its brothers. The advantage of this is conceptual
simplicity, and the fact that the population will never explode. Also,
it is ideally suited for massively parallel machines where each node
handles one or a small number of replicas. The drawbacks are that 
typical implementations might not be completely free of bias, that 
it needs substantial memory space, and, if population control is 
frequent, it may lead to heavy communication.

An alternative is depth first implementation\cite{tarjan}. There one 
keeps only one replica at each moment, and puts markers at times when
one decides for branching. The second branch is persued only when the 
present branch is abandoned, either because the full length of the 
polymer is reached or because it is killed. This backtracking is 
implemented most easily by recursive function calls: Each addition 
of a particle involves a call of a subroutine, and branching into 
$k$ branches corresponds to $k$ calls immediately following each 
other. While this is very compact and elegant, the main problem is 
that it is not immediately obvious how to decide on when to kill or 
branch. The solution is given by the observation that the average 
weight is an unbiased estimator of the partition sum (for an 
elementary discussion see the appendix of Ref\cite{hng2003a}),
\be
   \hat{Z}_n = \frac{1}{M} \sum_{\alpha=1}^M W_n^{(\alpha)}\;,
\ee
where $\alpha = 1,2,\ldots$ denotes the trials. One kills (branches)
if the current weight is smaller (larger) by a given factor than 
the average weight of all previous configurations, i.e. than the 
estimated partition sum. The precise value of the factor is not 
important, values between 2 and 10 are usually good. After the 
first configurations the estimated $\hat{Z}_n$ is typically too small,
whence the algorithm tends to accept too many configurations and 
makes too many branchings. This might lead to explosions of the 
population size, and tricks to avoid this are discussed in 
Ref\cite{frauenkron,bastolla}.

As we said, we used in our work the simplest possible rule for 
killing: low weight configurations are always killed with 
probability 1/2, no matter how bad they are. There are much more 
sophisticated rules discussed in the literature\cite{liu}, but 
we found them not to be necessary. It simply does not matter much
if a bad configuration is kept for a few more iterations before it is 
finally killed. On the other hand, we found that the details of 
the branching process are crucial. In our first 
papers\cite{frauenkron,bastolla,frauenkron2} we made
{\it exact clones} in each branching event. At very low temperatures, 
where Boltzmann factors are large, ``good" configurations can have
very large weights and splitting them into two branches might not 
be sufficient. Thus we allowed also branching into 3 or more copies.
While this was efficient in keeping the weights within narrow bounds,
it was however not fully successful: Branching is of course done in
the hope that different copies will follow different paths, thus 
leading to more diversity in the sample than without branching.
While this was true at high temperatures, it was not quite true
at temperatures where the native state can be reached. There, 
it might happen that one path has a weight much larger than all 
others, and thus all branches will follow this single path, leading 
to a loss of diversity. The same problem is encountered in genetic
algorithms if one is too greedy.

The way out proposed in Ref\cite{hng2003a} is to make {\it different}
first steps for each branch. Thus one first estimates (maybe crudely)
the weight after the next adding of a monomer. Depending on this one 
decides whether one wants to kill, continue, or branch -- and into 
how many branches. Here one also checks that $k$ different steps
are indeed sterically possible, if one wants to make $k$ branches.
After that decision is taken, one selects one from all possible 
$k$-tuples of steps, and continues with these $k$ branches.
Details of this, including the general problem of selecting 
random {\it tuples} in Monte Carlo simulations, are given in 
Ref\cite{hng2003a}. 

The above strategy is straightforwardly implemented in lattice 
models where the number of a priori possible continuations (the number 
of ``candidates") at each step is given by the coordination number. 
To implement it for off-lattice models one first has to select a 
random list of candidates (this is done independently at each step),
before one selects tuples from this list\cite{perm97}. For polymers 
above and near the $\Theta$ collapse temperature the optimal number 
of candidates is $\approx 4$\cite{perm97}. For ground state searches 
it turned out to be much larger, typically 20 to 50\cite{hvg-still}.

The algorithm with identical clones was called PERM (pruned-enriched
Rosenbluth method) in Ref\cite{perm97}, the version with branches forced 
to be different was called nPERM (new PERM) in Ref\cite{hng2003a}. A
further advantage of the latter is that it crosses smoothly over to 
exact enumeration, if the branching threshold is lowered to zero.
Thus one can e.g. make simulations where one follows all possible 
paths during the first steps, and makes random samplings only for 
the later steps. 

Results obtained with nPERM will be discussed in the next two 
sections. In Sec.3 I will treat lattice models, while off-lattice
models will be the subject of Sec.4.

\section{Lattice models}

\subsection{The HP - model}

The most popular lattice model for protein folding is the HP model 
of K. Dill and coworkers\cite{dill,lau-dill,chan-dill}. It assumes 
that hydrophobicity is the main driving force in folding, and that 
all amino acids can be classified into just two groups: Hydrophobic 
ones (``H") which avoid contacts with water and thus form the core
of a folded protein, and polar ones (``P") which want to be on the 
surface. This simplification is merged with the assumption that 
amino acids sit at the vertices of a regular lattice (square for 2-d,
simple cubic or FCC for 3-d), to form a class of extremely minimalistic
models. Proteins are modelled in this model by self avoiding walks
with attractive interactions between neighbouring non-bonded H-H pairs:
The energy between two hydrophobic monomers is -1 unit, while there 
is zero contact energy between P-P and H-P pairs.

The HP model has been justly criticized for being too much
simplified\cite{kaya-chan}, but it might nevertheless be very useful
as a test bed for folding algorithms. In particular this is true 
because exact ground states can be obtained by combinatorial 
methods\cite{yd93,yd95,will-backofen} for medium long chains, so one can 
verify in many cases whether the ground state is actually reached.
We should point out that these exact methods work {\it only} for 
the HP model and assume the existence of a well-formed hydrophobic 
core (in the absence of which they don't give wrong results but 
simply don't converge), while our Monte Carlo method is in principle 
a blind general purpose method.

There are a number of benchmark test cases which were discussed 
thoroughly in Ref\cite{hng2003a}. In the following we shall touch only 
some of them, and shall instead concentrate on examples which were 
overlooked in Ref\cite{hng2003a}.

{\bf (a)} A set of ten 48-mers on the SC lattice was studied in 
Ref\cite{yue95}. This was meant as a test for the ability of conventional
Monte Carlo search methods, and these methods failed in the majority
of cases. With old PERM we could already fold all ten, with new PERM
the CPU times were further cut down by 1 to 2 orders of magnitude 
to typically 10 seconds to 10 minutes on a fast PC\cite{hng2003a}.

{\bf (b)} A 85-mer 2-d HP sequence was given in Ref\cite{konig}, where it was
claimed to have $E_{\rm min}=-52$. Using a genetic algorithm, the authors
could find only conformations with $E\ge -47$. In Ref\cite{lw01}, using
a newly developed {\it evolutionary Monte Carlo} (EMC) method, the authors
found the putative ground state only when assuming large parts of its known
structure as constraints. Within ca. 1 minute on a fast PC, we found states
with $E = -53$.

\begin{figure}
\begin{center}
  \psfig{file=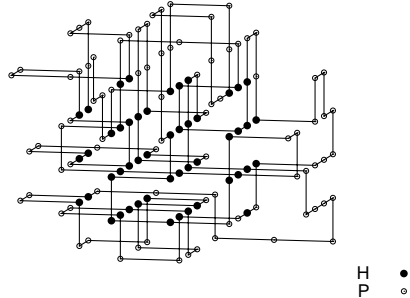,width=5.4cm,angle=270}
  \vglue - 6mm
  \caption{Configuration with $E=-80$ of a $N=136$ HP sequence modeling
  a staphylococcal nuclease fragment.}
\label{toma136}
  \vglue - 1mm
\end{center}
\end{figure}

{\bf (c)} Four 3-d HP sequences with lengths between $N=58$ and $N=136$ 
were proposed in Ref\cite{dfc93,lfd94} as models for actual proteins or 
protein fragments. Lowest energy states for these sequences were searched
in Ref\cite{tt96} using a newly developed and supposedly very
efficient algorithm. With nPERM, we now found states with lower energy 
after only few minutes CPU time, for all four chains. For
the longest one, the lowest energy found by us is 15 units lower than 
the supposed ground state energy found in Ref\cite{tt96}, but we are not 
sure that we did hit the true ground state. The chain is too long for the 
exact algorithm of Ref\cite{yd93,yd95}. But Fig.~1 shows that our lowest 
energy state has a rather compact hydrophobic core, so it could well be 
a ground state.

\begin{table}
\begin{center}
\caption{ Lowest energies of the 10 3-dimensional HP sequences
   of Unger {\it et al.}, see paragraph (d).} \label{table1}
\begin{tabular}{ccccc}
 sequence Nr. & Ref\cite{unger64} & Ref\cite{patton} & nPERM & lower bounds \\ \hline
    1  &  -27  &  -27   &     -32  &    -33 \\
    2  &  -29  &  -30   &     -38  &    -41 \\
    3  &  -35  &  -38   &     -45  &    -49 \\
    4  &  -34  &  -34   &     -42  &    -45 \\
    5  &  -32  &  -36   &     -43  &    -48 \\
    6  &  -29  &  -31   &     -35  &    -37 \\
    7  &  -20  &  -25   &     -28  &    -32 \\
    8  &  -29  &  -34   &     -38  &    -43 \\
    9  &  -32  &  -33   &     -41  &    -45 \\
   10  &  -24  &  -26   &     -31  &    -35 \\ \hline
\end{tabular}
\end{center}
\end{table}

{\bf (d)} Ten random 64-mers on the SC lattice were chosen as tests in
Ref\cite{unger64} for a genetic algorithm. This algorithm was criticized 
by Patton {\it et al.}\cite{patton} as being not optimal and too much 
influenced by Monte Carlo philosophy. Indeed, with a supposedly optimal 
genetic algorithm Patton {\it et al.} found in most cases lower energy 
states, by up to 5 units. Unfortunately no timings were given in 
Ref\cite{unger64,patton}. With nPERM we found in {\it all} cases even lower 
energies than Patton {\it et al.}, by up to 8 units (see Table 1). Our 
energies are rather close to lower bounds obtained by estimating the
hydrophobic core, so some of them might be real ground states. While 
reaching some of the lowest energies needed up to 8 hours on a 2 GHz
PC, others were reached within seconds (in which case it was checked 
that no lower energies were reached even after a day of CPU time). 
All energies reached in Ref\cite{patton} would have been reached within
less than half a second.

{\bf (e)} In these simulations we have tried to let the chains grow from 
either end. Sometimes growing from one end is more successful than growing 
from the other. This is even more pronounced in some other sequences, 
one example of which is shown in Fig.2. This sequence was devised by 
Yue and Dill\cite{yd95} such that the $\alpha/\beta$ barrel shown in 
Fig.2 is provably the ground state. When we tried to find this configuration
by starting with the monomer \#1, nPERM did not succeed, because one 
encounters repeatedly structures which are very unstable during the growth 
phase. They would become stabilized later when more monomers are added, 
but before that could happen the configuration is killed as being not 
promising. What is lacking in PERM (as in any growth algorithm known to
us) is an efficient way to look sufficiently far ahead so that bottlenecks formed 
by configurations looking bad to a short-sighted eye can be overcome.
Of course one can make look-aheads where one scans all configurations
up to a few (typically 2 to 5) steps ahead\cite{zhang02}, but we 
found this in general to be too time consuming for being efficient.

\begin{figure}
 \begin{center}
   \psfig{file=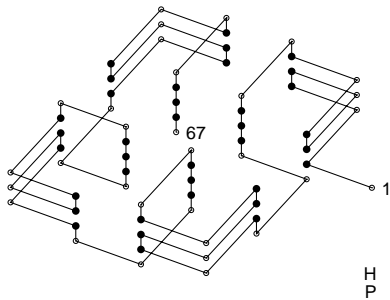,width=5.4cm,angle=270}
   \vglue - 4mm
   \caption{Ground state configuration ($E=-56$) of the $N=67$ HP sequence
    given by Yue and Dill. It forms a structure resembling an $\alpha/\beta$ 
    barrel. When starting at monomer \#67 ($\beta$ sheet end), nPERM could 
    find it easily, but not when starting from monomer \#1.}
\label{dill67}
  \vglue - 1mm
\end{center}
\end{figure}

{\bf (f)} Some of the lowest energy states shown in Table 1 were obtained
by a further trick. When placing a polar (P) monomer next to an H, one 
might block a site which would be needed during later stages of the growth.
Thus it is advantageous to disfavour the contact between H and P, even 
though such contacts are energetically neutral. Formally this is done 
by adding a repulsive P-H potential (of $\approx 0.2$ units) which 
is included in the Boltzmann factors but which is of course not included 
in the final energy count. This means that folding becomes easier if 
one keeps the chain {\it less compact} during the growth. It seems to 
disagree with claims to the contrary, albeit made in a different context,
in Ref\cite{kd01}.

{\bf (g)} It is argued in Ref\cite{will-backofen} that the FCC lattice is 
more typical for the internal structure of real proteins than the SC 
lattice. Indeed, it seems that it leads more easily to a hydrophobic core 
construction, otherwise it would be hard to understand why the code 
of Ref\cite{will-backofen} is so much more efficient than that of Ref\cite{yd93,yd95}.
It finds for most randomly chosen HP sequences of length $N=100$ the 
ground state(s), while the SC code of Ref\cite{yd93,yd95} in general breaks
down at $N\approx 80$. The same is true for nPERM. We found that we could 
reach the true ground states for several randomly chosen sequences of 
length $N=80$ on the FCC lattice\cite{will-server}, while chains of 
the same length on the SC lattice would have posed serious problems.
 
\subsection{Other models}

In Ref\cite{bastolla} we had used (old) PERM to find the ground state of 
a chain with two types of monomers which were however not of the H-P
type\cite{tp92}. This model is interesting because it shows an unexpected
transition from a 4-helix bundle to a $\beta$-sheet dominated structure
at very low temperatures. Later the same ground state  was found also 
in Ref\cite{chikenji} using a different Monte Carlo method. With nPERM 
the ground state is found an order of magnitude faster. 

Models with more than 2 types of monomers were studied also in 
\cite{bastolla}, with no particular difficulties. Applying nPERM to 
the model of Ref\cite{kaya-chan} which has 5 types of monomers gave very
fast convergence (which should not surprise in view of the very small
length of this chain), and also to a very stable estimate of excited
states. Up to the tenth excited energy level the 
degeneracy factors were so close to integers that one could infer the 
exact numbers of excited states, although they went up to the hundreds
(unpublished).
The fact that PERM and nPERM are truly thermodynamic approaches 
and give thus rise to the full energy spectrum is indeed one of their
strong points.

\section{Off-lattice models}

There are much less results for off-lattice models in the literature
that can be used as benchmarks. Of course one could use realistic 
small proteins with one of the standard force fields, but the results
would be hard to compare with other algorithms. Thus we preferred simple
toy models. There is essentially one such toy model in the literature,
by Stillinger {\it et al.}\cite{sh95}. This is a two-dimensional model
with modified Lennard-Jones potentials, fixed bond lengths, and a weak 
bond angle stiffness term. It contains also two types of monomers 
mimicking hydrophobic and polar amino acids, but this time they are 
called ``A" (hydrophobic) and ``B" (polar). In Ref\cite{sh95} the authors 
studied only Fibonacci sequences where the chain length is a Fibonacci
number $F_i$, and there are $F_{i-2}$ A monomers and $F_{i-1}$ B's.

\begin{figure}[ht]
\begin{center}
$\begin{array}{c@{\hspace{0.1in}}c}
\multicolumn{1}{l}{\mbox{\tiny N = 13}} &
        \multicolumn{1}{l}{\mbox{\tiny N = 21}} \\ [-0.53cm]\\
\epsfig{file=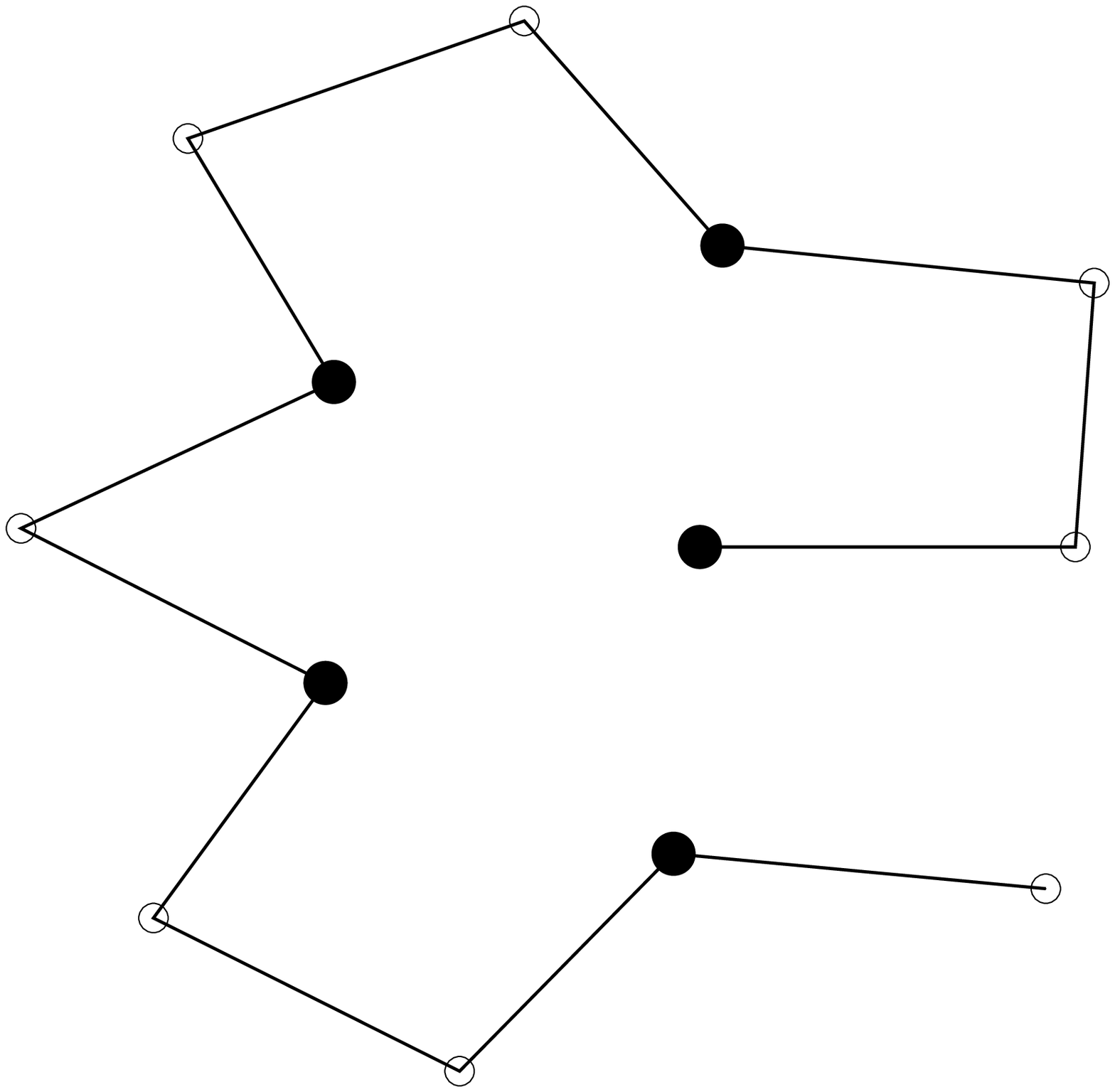, width=3.0cm, angle=270} &
\epsfig{file=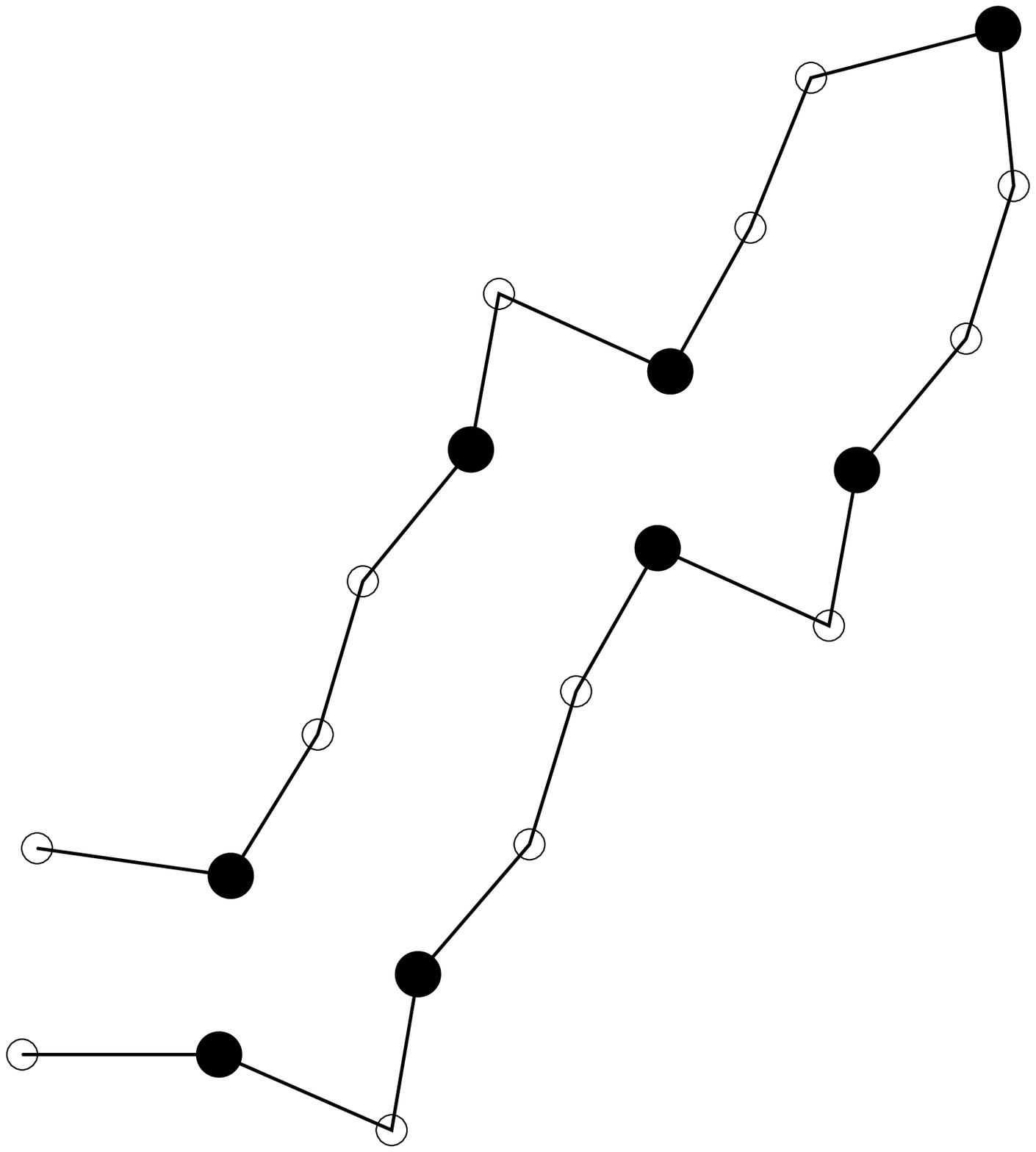, width=3.0cm, angle=270} \\
\multicolumn{1}{l}{\mbox{\tiny N = 34}} &
        \multicolumn{1}{l}{\mbox{\tiny N = 55}} \\ [-0.53cm]\\
\epsfig{file=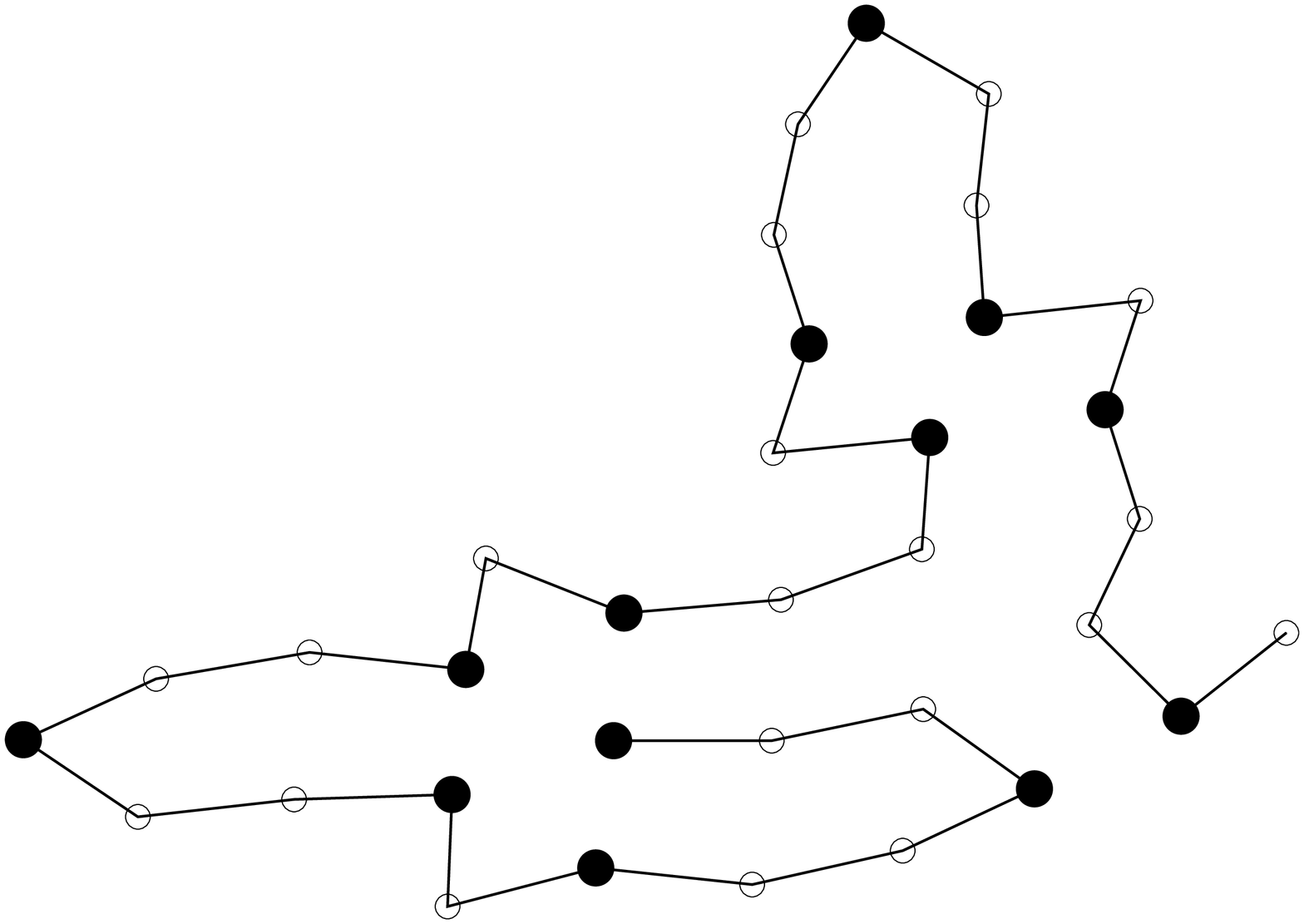, width=3.0cm, angle=270} &
\epsfig{file=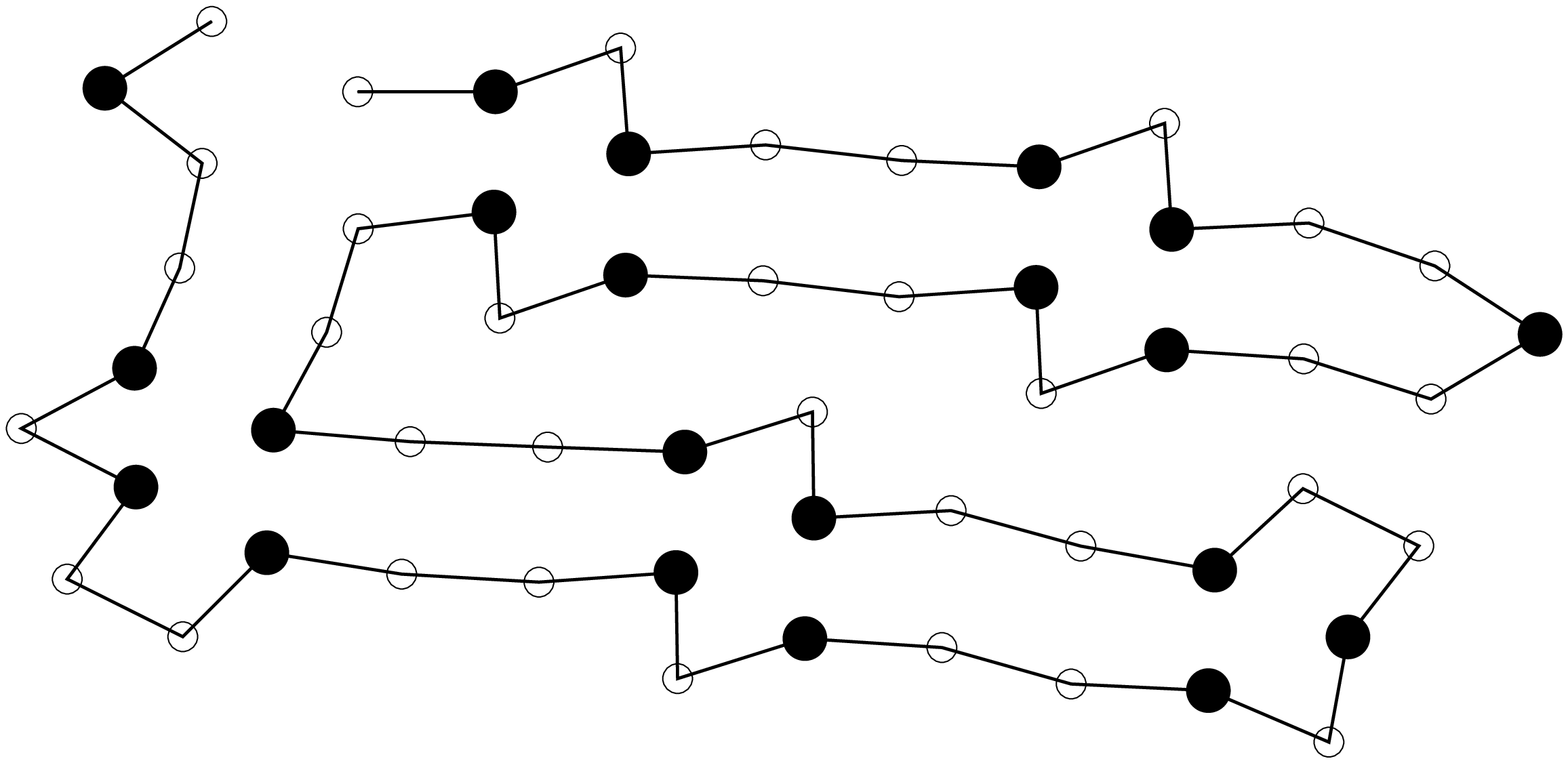, width=3.0cm, angle=270} \\
[0.4cm]
\end{array}$
\caption{ Putative 2-d ground states of Fibonacci sequences. Full dots 
   indicate hydrophobic monomers.}
\label{fig-2d}
\end{center}
\end{figure}

\begin{figure}[h]
\begin{center}
$\begin{array}{c@{\hspace{0.1in}}c}
\multicolumn{1}{l}{\mbox{\tiny N = 55}} & \multicolumn{1}{l}{} \\ [-0.53cm]
\epsfig{file=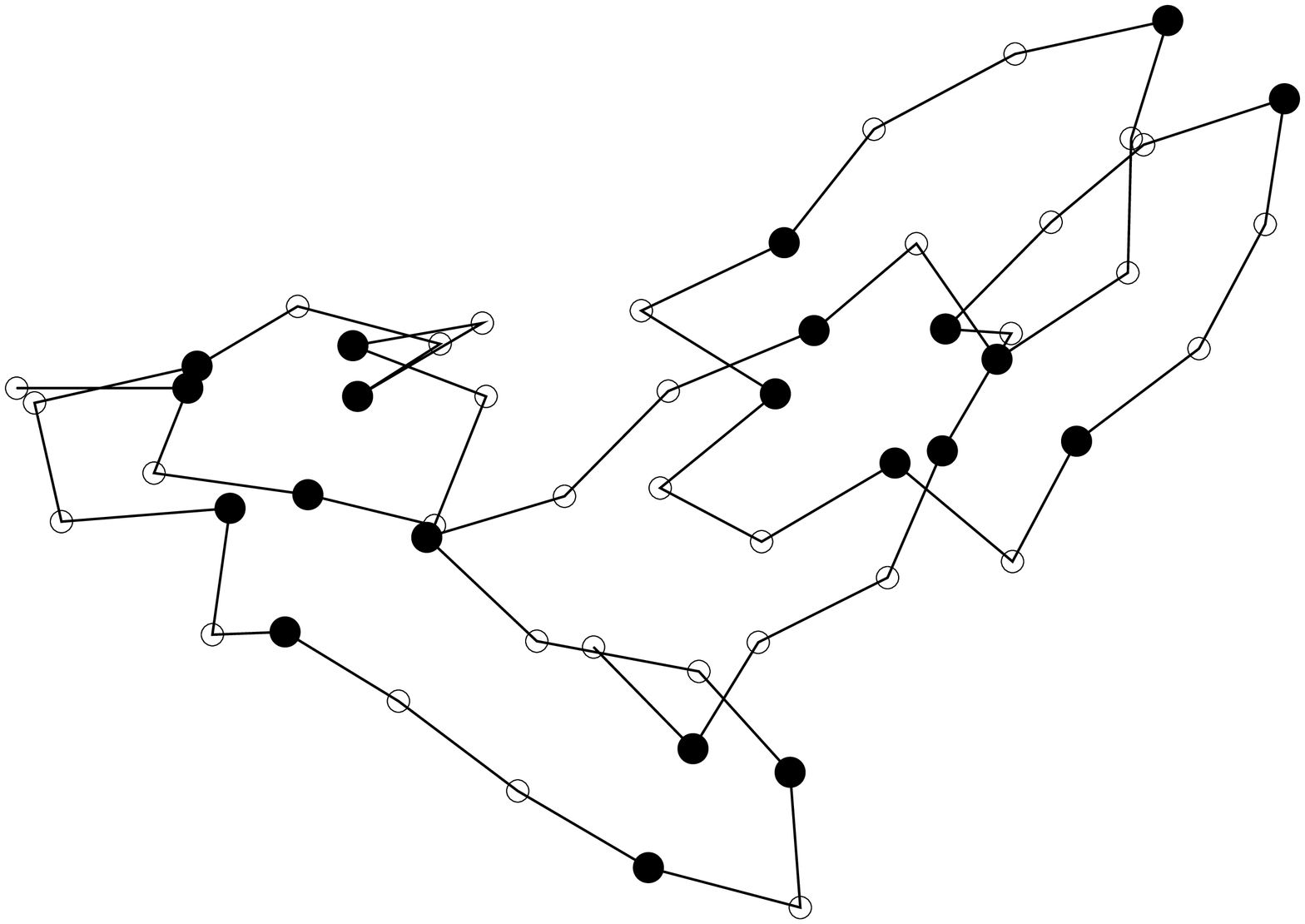, width=3.5cm, angle=270} &
\epsfig{file=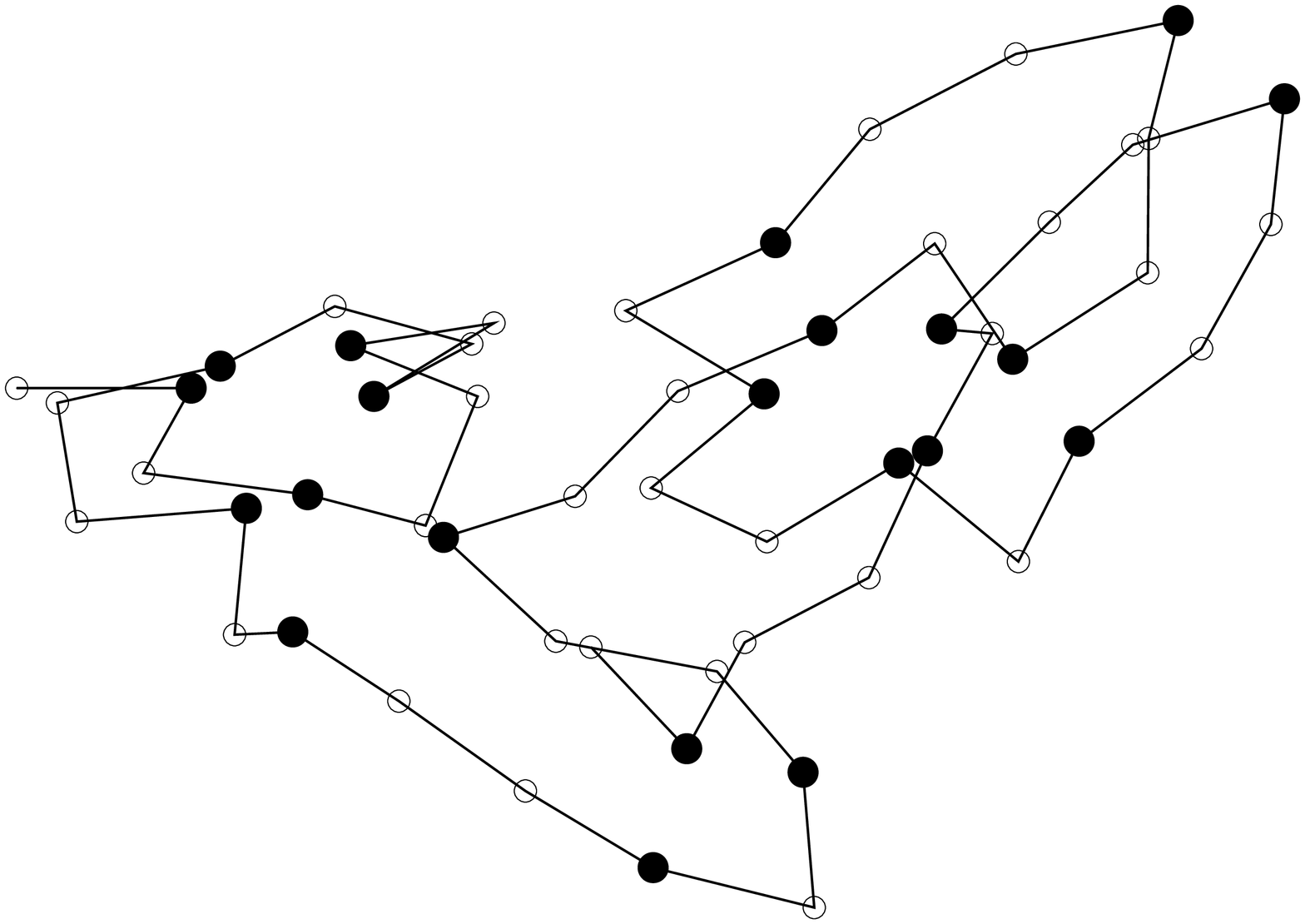, width=3.5cm, angle=270} \\
\end{array}$
\caption{ Stereographic view of the putative ground state of the $3-d$ Fibonacci
sequence with $N=55$. Again, $A$ monomers are shown as filled circles.}
\label{fig-3d}
\end{center}
\end{figure}

Putative ground states were found in Ref\cite{hvg-still} by first using nPERM 
to come close to a potential minimum, and then using conjugate gradient 
descent to reach the minimum itself. Results are depicted in Fig. 3. 
In all cases the energies are smaller than those reported in 
Ref\cite{sh95}, e.g. for $N=55$ the configuration shown in Fig. 3 has 
$E=-18.515$, while the supposed ground state given in Ref\cite{sh95} 
had $E=-14.41$. For the longer chains also the topology is different.
A very striking feature is that only the shortest chain (with $N=13$)
has a connected hydrophobic core. Thus this is a very bad model for 
real proteins. The ultimate reason for this is the absence of longer 
connected stretches of hydrophobic monomers along the chain. In a
Fibonacci chain as defined in Ref\cite{sh95}, all A monomers are flanked 
by two B monomers, which prevents the formation of large hydrophobic
cores.

The same model was used in Ref\cite{hvg-still} for 3-d structures. The 
potential energy was kept identical to the above, although a different
bond angle term favouring a more Ramachandran-like torsion angle pattern
would have been more physical. But the Fibonacci structure of the AB
sequences prevented also in 3-d the formation of a single core (see Fig. 3), 
and thus the model is not very realistic anyhow. We present it here 
merely for future benchmarks.

\section{Conclusion}

In this review we have not attempted to study very realistic protein
models. Rather, we wanted to present a class of general purpose 
algorithms which perform extremely well for simple lattice polymers, 
and which therefore have a potential of being useful also in more complex
applications. When applied to lattice models, nPERM seems to be superior
than genetic algorithms and all known Markov chain (Metropolis-type)
Monte Carlo methods, and is matched only by exact methods which are 
however very specifically taylored to these models. For off-lattice
models the same algorithm fares quite well, even if it might not be 
better than modern Markov chain methods. We have not yet tested it 
for proteins surrounded by solvents, where previous Monte Carlo methods
seem (still) inferior to molecular dynamics. As it is formulated now,
we expect that PERM and nPERM would share the same problems as 
Metropolis-type Monte Carlo. But we cannot exclude that future
progress might lead to substantial improvements. We just want to 
mention that a first step has been made in Ref\cite{bach-janke,prellberg},
where PERM was combined with the multicanonical (umbrella sampling)
concept, leading to very promising results. In the long run, we expect
that a mix of different Monte Carlo strategies will lead finally to 
a substantially improved understanding of protein folding.

Acknowledgements: I want to thank my collaborators U. Bastolla, 
H. Frauenkron, E. Gerstner, S. M. Causo, B. Coluzzi, W. Nadler, 
H.-P. Hsu, and V. Mehra for the very fruitful and pleasant joint 
efforts. I also want to thank B. Berg and U. Hansmann for many
useful discussions.

\end{document}